\documentclass[sigplan,nonacm]{acmart}
\AtBeginDocument{%
  }

\setcopyright{acmlicensed}
\copyrightyear{2018}
\acmYear{2018}
\acmDOI{XXXXXXX.XXXXXXX}
\acmConference[Conference acronym 'XX]{Make sure to enter the correct
  conference title from your rights confirmation email}{June 03--05,
  2018}{Woodstock, NY}
\acmISBN{978-1-4503-XXXX-X/2018/06}

\usepackage{tikz}
\usepackage{amsmath}

\usepackage[linesnumbered,ruled,vlined]{algorithm2e}

\usepackage{filecontents}

\usepackage{pifont}

\usepackage{algpseudocode}

\usepackage{booktabs}

\usepackage{enumitem}

\usepackage{mdframed}

\newmdenv[
  skipabove=3pt,
  skipbelow=1pt,
  backgroundcolor=black!3,
]{insightbox}

\newmdenv[
  backgroundcolor=black!8,
  linewidth=0.7pt,
  linecolor=black!35,
  leftline=true,rightline=false,topline=false,bottomline=false,
]{keybox}

\newcommand{\bulletpara}[1]{%
  \par
  \textbullet\ \textbf{#1}%
}
\SetAlFnt{\small}   

\SetKwInput{KwIn}{Input}       
\SetKwInput{KwParam}{Param}    
\SetKwInput{KwState}{State}    

\begin{document}

\title{TIDAL: Recovering Temporal Phase for Cloud Block Storage Placement from LLM-Derived Semantics}

\author{Difan Tan}
\email{difan_tan@hust.edu.cn}
\orcid{0009-0003-7713-2378}
\affiliation{%
  \institution{Huazhong University of Science and Technology}
  \country{}
}

\author{Changlin Wan}
\affiliation{%
  \institution{Huazhong University of Science and Technology}
  \country{}
}

\author{Jiawen Liu}
\affiliation{%
  \institution{The Hong Kong Polytechnic University}
  \country{}
}

\author{Hua Wang}
\authornote{Corresponding author.}
\email{hwang@hust.edu.cn}
\affiliation{%
  \institution{Huazhong University of Science and Technology}
  \country{}
}

\author{Ke Zhou}
\affiliation{%
  \institution{Huazhong University of Science and Technology}
  \country{}
}

\renewcommand{\shortauthors}{Trovato et al.}


\begin{abstract}

Cloud Virtual Disk (CVD) placement in Cloud Block Storage (CBS) is critical for resource efficiency and performance isolation. Existing schemes prioritize spatial load balancing by dispersing disks across pods based on configuration-derived load estimates. However, overload risk in CBS is fundamentally temporal. Even when average load is balanced, pods can still suffer transient congestion when the peaks of co-located disks align in time. Achieving complementary placement, which co-locates CVDs with offset peaks, is hard at provisioning time because new disks have no history from which to infer temporal phase.

We present \textbf{TIDAL}, a CVD placement framework that recovers phase-aware signals for cold-start placement from an underused source: tenant-provided names and identifiers in provisioning metadata. TIDAL first uses LLMs to recover application semantics from noisy metadata such as project, VM, and disk names. It then translates these semantics into phase-aware temporal signals to guide complementary placement. To satisfy control-plane constraints, TIDAL adopts an offline-to-online design with teacher-student distillation, regex-based filtering, and prefix-aware caching, enabling CPU-only inference with millisecond-level latency. Evaluations driven by production traces show that TIDAL reduces overload frequency by 79.1\% and P95 overload duration by 73.7\% compared with the strongest baselines.

\end{abstract}



\maketitle

\section{Introduction}\label{sec:intro}
Cloud Block Storage (CBS) is a core building block of modern clouds, providing persistent block devices for Cloud Virtual Machines (CVMs) across providers such as AWS~\cite{Amazon}, Microsoft Azure~\cite{Azure}, and Alibaba Cloud~\cite{EBS_evolut}. In CBS, tenants provision Cloud Virtual Disks (CVDs)~\cite{VD_origin, CBS_VD_load, SCDA, CBS_load_heyhey, vd_load} to scale storage resources, which are placed onto backend storage clusters partitioned into \emph{pods}. The placement of a newly created CVD---that is, deciding which pod hosts it at provisioning time---directly affects both performance isolation and hardware efficiency.

\begin{figure}[t]
\centerline{\includegraphics[width=0.48\textwidth]{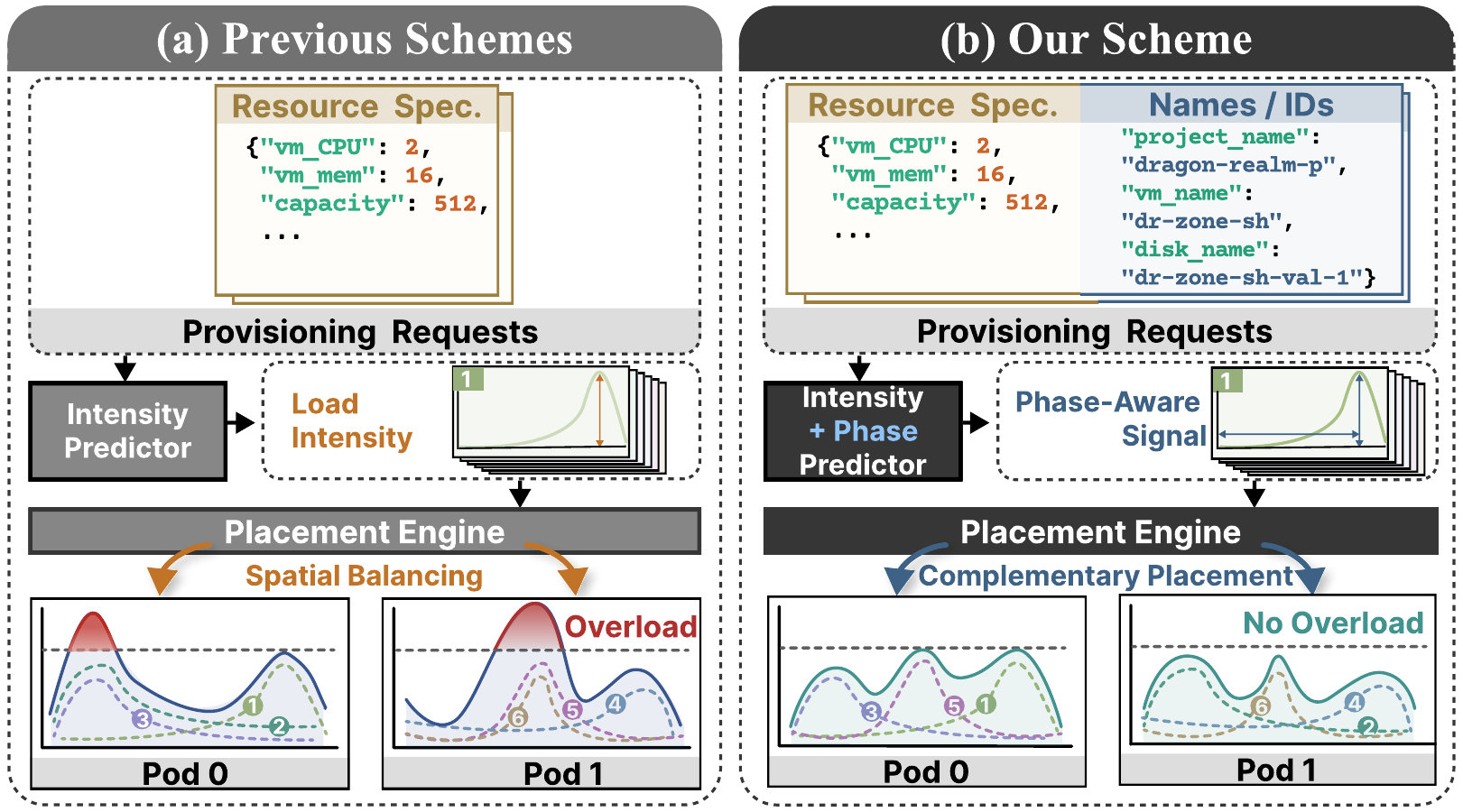}}
\caption{Comparison between existing CVD placement schemes (a) and TIDAL (b).}
\label{intro}
\end{figure}

A core objective of CVD placement is to maintain load balance across pods. This matters because severe imbalance can lead to \emph{overload}~\cite{overload}: scenarios where a pod's aggregate throughput demand exceeds its hardware bandwidth, triggering I/O throttling and SLA violations~\cite{overload_defcon, CBS_load_heyhey}. Existing schemes~\cite{SCDA,TELA} therefore focus on \emph{spatial balancing}. 
As Figure~\ref{intro}(a) illustrates, they estimate a disk's expected load intensity from resource specifications in the provisioning request and scatter disks to equalize utilization across pods. In the example, disks \ding{182}--\ding{184} are placed in Pod~0 and \ding{185}--\ding{187} in Pod~1, yielding nearly balanced average load. Yet overload still occurs because disks placed in the same pod peak at similar times. Avoiding such peak coincidence requires \emph{complementary placement}~\cite{vm_complement}: placing a new disk into a pod whose existing workloads tend to peak at different times. Doing so requires knowing when the new disk is likely to peak relative to existing workloads, i.e., its \emph{temporal phase}.

The difficulty is that this temporal phase is needed exactly when it is least observable. Placement decisions are made at provisioning time, when a new disk has no historical trace. Existing prediction-based schemes~\cite{TELA,SCDA} rely on resource specifications in the provisioning request. These fields correlate with load intensity, but provide little information about temporal phase. Even disks with the same configuration can exhibit very different intra-day behavior; for example, a database server and a gaming server may both be provisioned as ``512\,GB SSD'' yet follow very different diurnal rhythms. Without runtime history, the scheduler is therefore blind to whether a new disk is likely to reinforce existing peaks or fill existing valleys.

This raises a natural question: how can a new disk's temporal phase be inferred at provisioning time? Our key insight is that provisioning metadata also contains tenant-provided names and identifiers---such as \verb|project_name|, \verb|vm_name|, and \verb|disk_name|---that implicitly encode application semantics. These fields are largely underutilized in existing placement schemes because they are free-form and difficult to interpret reliably. The application semantics they carry, however, are often associated with recurring temporal patterns (detailed in \S\ref{sec:ideavalidate}). For example, gaming services may peak at night, while office applications are more active during the day. This suggests a natural idea as shown in Figure~\ref{intro}(b): infer a disk's likely temporal phase from its metadata, and use that phase signal to guide complementary placement.

Realizing this metadata-to-placement pipeline in practice requires solving three challenges. \textbf{First}, tenant-provided identifiers are noisy, open-vocabulary, and often non-literal, making semantic recovery difficult; rule-based or keyword-based methods~\cite{text_class} break easily on internal codenames and tenant-specific naming conventions. \textbf{Second}, the recovered semantics must be translated into a schedulable phase-aware signal that can be combined with pod-level load state for complementary placement. \textbf{Third}, the entire pipeline must fit within the control plane latency budget. Directly invoking heavyweight semantic models on the critical path would incur unacceptable latency~\cite{IC-cache,SpecInfer} and expensive GPU serving cost, increasing total cost of ownership (TCO)~\cite{vllm,severlessllm}.

To address these challenges, we present \textbf{TIDAL}, an end-to-end placement framework that realizes the design in Figure~\ref{intro}(b). TIDAL first uses large language models (LLMs)~\cite{llm_gpt4,llm_gemini,llm_llama2,llm_deepseek,llm_qwen} to recover application semantics from noisy provisioning metadata. It then maps the recovered semantics to representative temporal patterns and combines them with predicted load intensity to guide complementary placement. To meet control-plane latency constraints, TIDAL adopts an offline-to-online design: a heavyweight teacher LLM generates supervision offline, a lightweight CPU-efficient student model performs online inference, a regex-based filter bypasses obviously non-semantic requests, and a prefix-aware semantic cache short-circuits redundant runtime inferences. Together, these techniques enable sub-10\,ms placement decisions while preserving the benefits of LLM-based semantic understanding.

This paper makes the following contributions:

\begin{itemize}[itemsep=0pt]
    \item We formulate temporal phase recovery for newly provisioned disks as a missing capability in CBS placement, and show that resource specifications alone are insufficient for this task.
    \item We present TIDAL, an end-to-end placement architecture that turns LLM-derived application semantics into actionable temporal-phase signals for complementary CVD placement without historical traces.
    \item We design a practical control-plane realization via teacher--student distillation, regex-based filtering, and prefix-aware caching, reducing average placement latency to 7.03\,ms on a CPU-only server.
    \item We evaluate TIDAL on real production traces. TIDAL reduces overload frequency by 79.1\% and P95 overload duration by 73.7\% compared to the best existing baseline.
\end{itemize}

\section{Background}\label{sec:bg}
\subsection{Cloud Virtual Disk Placement}
\label{sec:cvdp}
\begin{figure}[t]
\centerline{\includegraphics[width=0.45\textwidth]{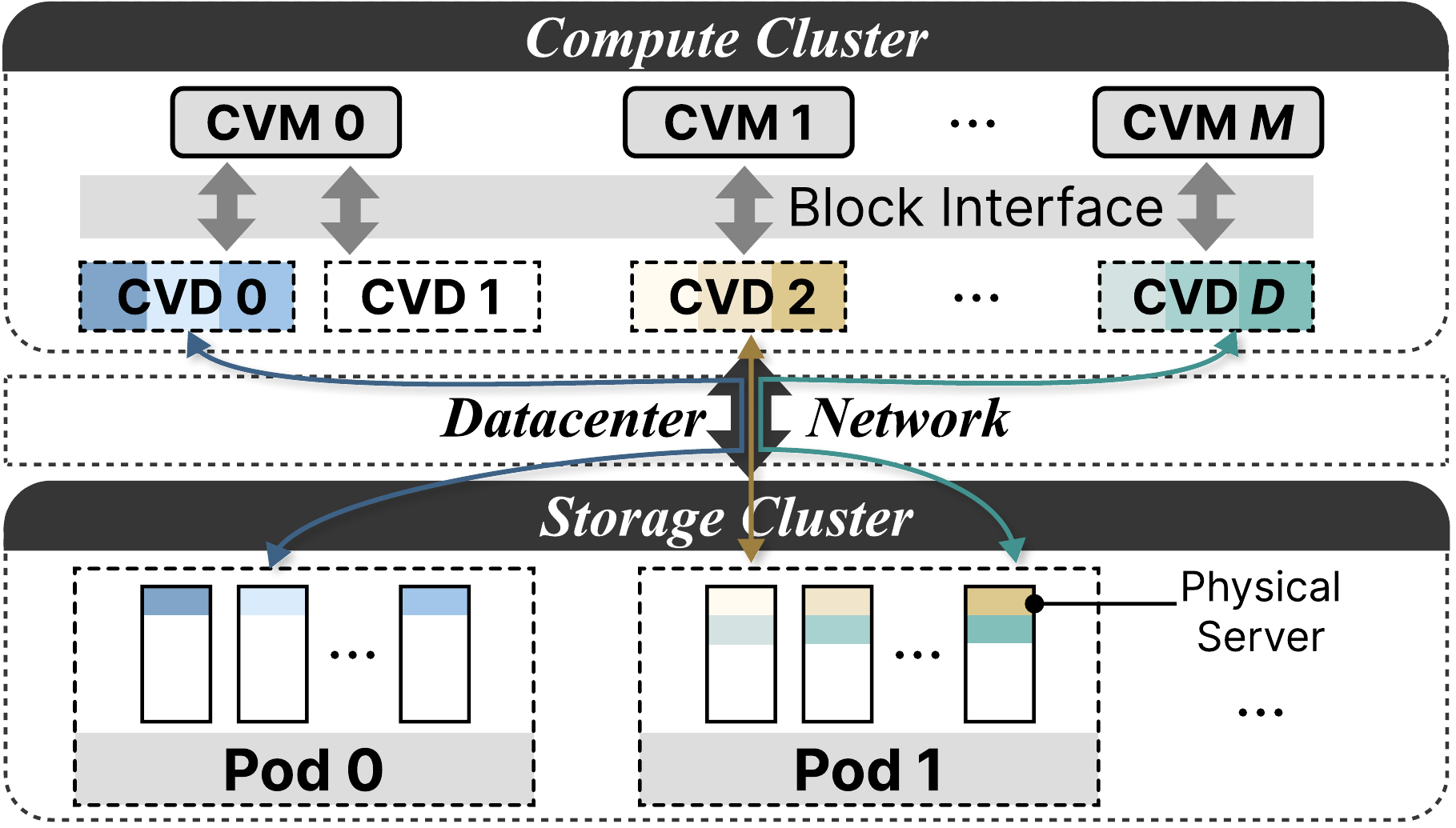}}
\caption{Cloud block storage architecture. CVMs access CVDs over the datacenter network; the storage cluster is partitioned into pods spanning multiple physical servers.}
\label{bg1}
\end{figure}

\noindent \textbf{CBS architecture.} Figure~\ref{bg1} shows a typical cloud block storage (CBS) architecture. In compute--storage disaggregated designs~\cite{EBS_DPU,EBS_recovery}, CVMs run in compute clusters while CVDs reside in separate storage clusters, and I/O traverses the datacenter network between them. For manageability and fault isolation~\cite{TELA}, the storage backend is partitioned into \emph{pods}, which are the unit of placement. Within a pod, a CVD may be striped across multiple servers for parallelism and intra-pod load balancing~\cite{EBS_DPU,EBS_evolut,CBS_load_heyhey}.

\noindent \textbf{Placement decision and balancing limits.} CVD placement determines the target pod of a newly provisioned disk and therefore sets the initial load distribution across pods. Although post-placement migration can rebalance load~\cite{VDMig,VD_migration}, it is expensive and reactive: copying persistent data consumes storage and network bandwidth, and corrective action occurs only after imbalance has already appeared. Many schemes therefore focus on optimizing initial placement, typically through \emph{spatial balancing}. They estimate a disk's load intensity from resource specifications in the provisioning request and scatter disks to equalize utilization across pods~\cite{SCDA,TELA}. However, spatial balancing alone is insufficient. Even when average load is balanced across pods, synchronized peaks within a pod can still push aggregate throughput beyond pod capacity, triggering overload, I/O throttling, inflated tail latency, and potentially SLA violations~\cite{overload,overload_defcon,CBS_load_heyhey}. This motivates \emph{temporal balancing}, which in our setting is achieved through \emph{complementary placement}: placing a new CVD into a pod whose existing load profile offers complementary valleys rather than coincident peaks.

\subsection{Temporal Characteristics of CBS Workloads}\label{2b}
\label{sec:workload}
To quantify the opportunity for temporal balancing, we analyze production traces from CloudProvider-A (anonymized): 32{,}000 CVDs sampled from 16 clusters over 14 consecutive days. This analysis reveals two key temporal properties of CBS workloads: stable intra-day phase behavior and diversified peak timing across disks.

\begin{figure}[t]
\centerline{\includegraphics[width=0.48\textwidth]{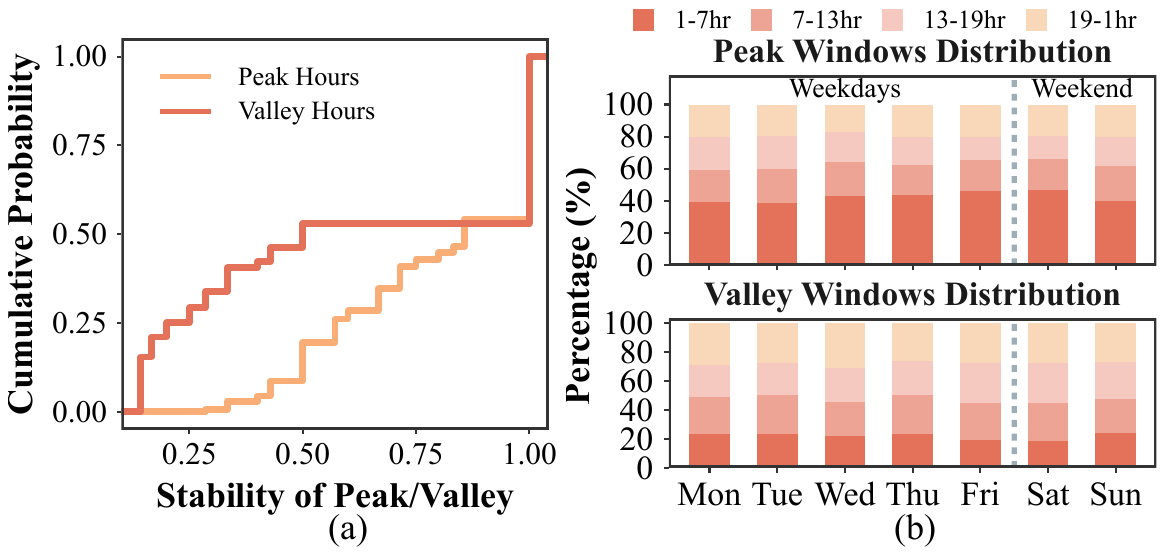}}
\caption{Temporal characteristics of CVD workloads. 
(a) CDF of peak/valley stability. 
(b) Distribution of peak/valley windows over weekdays and weekends.}
\label{bg2}
\end{figure}

\textbf{First, many CVDs exhibit stable intra-day phases.} Prior studies on AliCloud and Tencent Cloud report coarse-grained periodic patterns~\cite{CBS_load_zou, TELA}. We refine this observation by partitioning each day into four six-hour windows. We then examine whether the windows containing a CVD's daily peaks and valleys remain stable across days. For each CVD, we define \emph{stability} as the fraction of days on which its peak (or valley) falls into its most frequent window. Figure~\ref{bg2}(a) plots the resulting CDFs. The distributions are skewed toward higher stability values, indicating that many CVDs show consistent peak and valley timing across days rather than random intra-day phase shifts.

\textbf{Second, different CVDs often exhibit offset peak and valley windows.} Prior analyses of AliCloud and Tencent Cloud workloads reported balanced day--night utilization~\cite{CBS_load_Li,CBS_load_zou}, suggesting coarse-grained complementarity. We examine this at finer granularity using the same four-window partition. For each CVD, we define its \emph{peak window} as the window in which its daily peaks most frequently occur; we define its \emph{valley window} analogously. Figure~\ref{bg2}(b) shows that peak and valley windows are distributed across all four windows rather than concentrated in a single period. This is consistent with findings on Azure CVM workloads~\cite{coach}, where VM-level loads also exhibit diversified peak times. 

\begin{insightbox}
\textbf{Takeaway 1:} Many CVDs exhibit stable intra-day phase behavior, and different CVDs often peak at different times. Together, these properties create real opportunity to smooth aggregate pod load through complementary placement.
\end{insightbox}

\subsection{The Cold-Start Phase Gap}
\label{sec:limitation}
These observations point to an opportunity: use complementary placement to assign a new CVD to a pod whose existing workloads offset the disk's future peaks. The difficulty is that a newly created CVD has no runtime trace, so its temporal phase cannot be directly observed at placement time.

Existing predictive schemes for resource management and scheduling do not close this gap for two reasons. First, many predictive schedulers rely on runtime history to forecast future dynamics~\cite{microserv_madu, VM_placement_TAPAS,vm_complement,VM_placement_rescen,blockflex}. This is inapplicable at creation time with no history. Second, history-free schemes typically rely on static features available at decision time~\cite{lava,Utilize_BYOM,power_pred,TELA,SCDA,ML_job_rubick}. These features describe resource specifications, deployment context, or other categorical attributes available at decision time. They can help estimate \emph{how much} load a new disk may generate, but not \emph{when} that load is likely to occur. For example, the same ``512 GB SSD'' configuration may correspond to an online gaming service with evening peaks or an office automation workload concentrated in business hours. 

We refer to this mismatch as the \emph{cold-start phase gap}: complementary placement requires phase-aware signals at provisioning time, but the information available at creation time does not provide them.

\begin{insightbox}
\textbf{Takeaway 2:} Breaking the cold-start phase gap requires a new source of provisioning-time information: one that is available before the disk begins serving I/O, yet still carries clues about its temporal phase.
\end{insightbox}


\section{The Semantic Opportunity and Practical Challenges}\label{sec:opportunity}
\subsection{Metadata as a Phase Proxy}\label{sec:ideavalidate}

Section~\ref{sec:limitation} shows that complementary placement requires phase-aware signals at provisioning time, when a new CVD has no runtime history. Our key observation is that provisioning metadata often contains tenant-provided identifiers, such as \verb|project_name|, \verb|vm_name|, and \verb|disk_name|, that implicitly encode application semantics. Although these fields are created for operational convenience rather than scheduling, they often reveal the application, service role, or business context to which the disk belongs---for example, whether a disk serves an office system, an e-commerce backend, or an online game. Because application semantics are often associated with recurring diurnal phases, these identifiers can provide the missing phase clue for complementary placement at decision time.

We next test whether metadata-derived semantics are sufficiently prevalent and temporally consistent to support complementary placement in practice. Using the same production data as in \S\ref{sec:workload}, we recover application semantics from provisioning metadata (the recovery mechanism is detailed in \S~\ref{sec:recovering}) and examine whether they provide a useful phase-aware signal for complementary placement.

\begin{table}[t]
\footnotesize
\centering
\setlength{\tabcolsep}{3pt}
\caption{Recovered application labels over CVDs.}
\renewcommand{\arraystretch}{0.95}
\label{tab:llm-coverage}
\begin{tabular}{l r @{\hskip 6pt} l r}
\toprule
\multicolumn{4}{c}{Application type} \\
\cmidrule(lr){1-4}
Type & \% & Type & \% \\
\midrule
Infra-node               & 9.52 & Corp-website    & 2.10 \\
Database                 &  8.47 & Education              & 1.99 \\
Gaming                     &  6.54 & Finance/payment                & 1.35 \\
Infra-message-queue      &  6.42 & Infra-jumpbox          & 1.13 \\
Dev-test-env             &  6.21 & Media/news             & 1.02 \\
Media/video/streaming    &  5.56 & Infra-cache            & 0.80 \\
Compute/simulation       &  4.99 & Ecommerce/retail              & 0.65 \\
Infra-logging/monitoring &  3.98 & Community              & 0.51 \\
Data-collection &  3.53 & Travel                 & 0.48 \\
Office-system            &  3.39 & Gov-public-service     & 0.37 \\
Generic-autoscaling      &  2.89 & Logistics/mobility     & 0.26 \\
Infra-coordination       &  2.59 & Delivery & 0.14 \\
AI/ML                    &  2.37 & Infra-cloud-function   & 0.08 \\
\cmidrule(lr){3-4}
IoT-SaaS-platform        &  2.34 & \textbf{\textit{Unknown}}       & 20.29 \\
\bottomrule
\end{tabular}
\end{table}

Table~\ref{tab:llm-coverage} summarizes the recovered application types. The offline recovery pipeline categorizes 79.71\% of disks, while 20.29\% remain Unknown\footnote{The semantic distribution in Table~\ref{tab:llm-coverage} is computed over our random sample. It reflects only the composition of this sample for research purposes, rather than the revenue structure of CloudProvider-A.}. Most Unknown cases are opaque machine-generated identifiers, such as UUID-like or hash-like names, that contain little recoverable semantic content. Overall, most provisioning requests contain semantically informative metadata.

\begin{figure}[t]
\centerline{\includegraphics[width=0.45\textwidth]{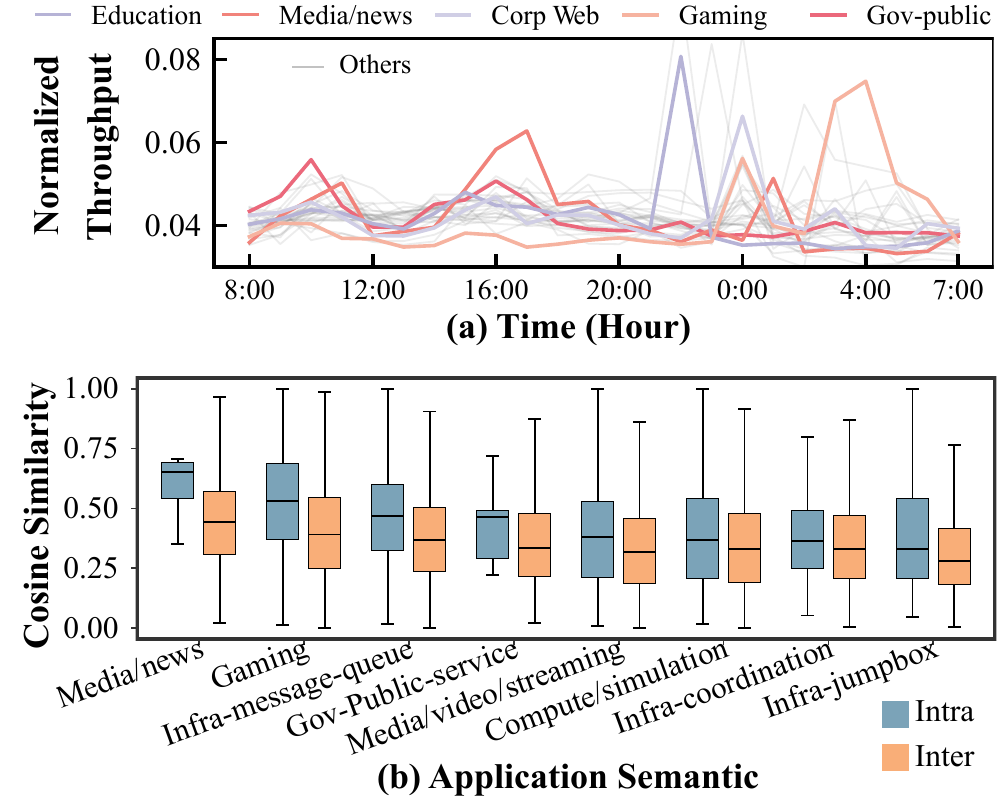}}
\caption{Category-wise I/O characteristics. (a) Representative normalized daily throughput profiles for recovered application classes. (b) Intra-class and inter-class cosine similarity of daily profiles grouped by the recovered semantic labels.}

\label{bg3}
\end{figure}

The recovered semantics are also temporally meaningful.
Figure~\ref{bg3}(a) shows representative daily profiles for several application classes, illustrating that different semantic groups can exhibit distinct intra-day rhythms. For example, gaming-related workloads peak during late-night hours\footnote{This pattern reflects disk I/O rather than front-end user activity. Late-night peaks may arise from backend synchronization, batch updates, or maintenance operations.}, whereas media or corporate web workloads are more active earlier in the day. 
Figure~\ref{bg3}(b) quantifies this separation using intra-class and inter-class cosine similarity for eight representative classes. Specifically, we show the four classes with the highest intra-class similarity and the four with the lowest, ordered from high to low along the x-axis. Even for the least compact classes, disks with same semantic remain more similar to each other than to disks from other groups, indicating that the recovered labels capture meaningful temporal structure rather than arbitrary textual similarity.

\subsection{From Opportunity to System Challenges}
\label{sec:challenge}
The evidence above establishes the semantic opportunity, but not yet a deployable placement mechanism. Turning metadata-derived semantics into a deployable placement system requires solving three distinct challenges.

\bulletpara{Challenge 1 (C1): Recovering semantics from noisy metadata.}
Tenant-provided identifiers in provisioning metadata are noisy, open-vocabulary, and often non-literal: a name such as \verb|dragon-realm-node-1| may imply a gaming workload instead of a "dragon" or "realm" workload, yet this semantics is not recoverable from lexical rules alone. Prior systems that incorporate textual features~\cite{LLAMA,VDMig} rely on limited mechanisms such as token matching or manually engineered text features, which remain brittle under noisy free-form metadata. The first challenge is therefore to recover application semantics from identifiers that were never designed for machine interpretation.

\bulletpara{Challenge 2 (C2): Converting semantics into placement decisions.}
Even if application semantics can be recovered, they are not directly schedulable. A semantic label alone does not determine where a disk should be placed. The second challenge is therefore to translate recovered semantics into concrete phase-aware signals and combine those signals with pod-level load state to drive complementary placement.

\bulletpara{Challenge 3 (C3): Meeting control-plane efficiency constraints.}
Placement lies on the critical path of provisioning, where decisions must be made within a tight latency budget on CPU-only control-plane nodes, typically on the order of 100\,ms. The third challenge is therefore to realize the full semantic-recovery and placement pipeline under strict efficiency constraints in the online path.

TIDAL is designed around these three challenges: it recovers application semantics from noisy metadata, converts those semantics into actionable phase-aware signals, and realizes the resulting pipeline within the efficiency constraints of CBS provisioning. Section~\ref{sec:design} presents this design.
\section{TIDAL Design}
\label{sec:design}

\subsection{Overview}
\label{sec:overview}

\begin{figure}[t]
\centerline{\includegraphics[width=0.48\textwidth]{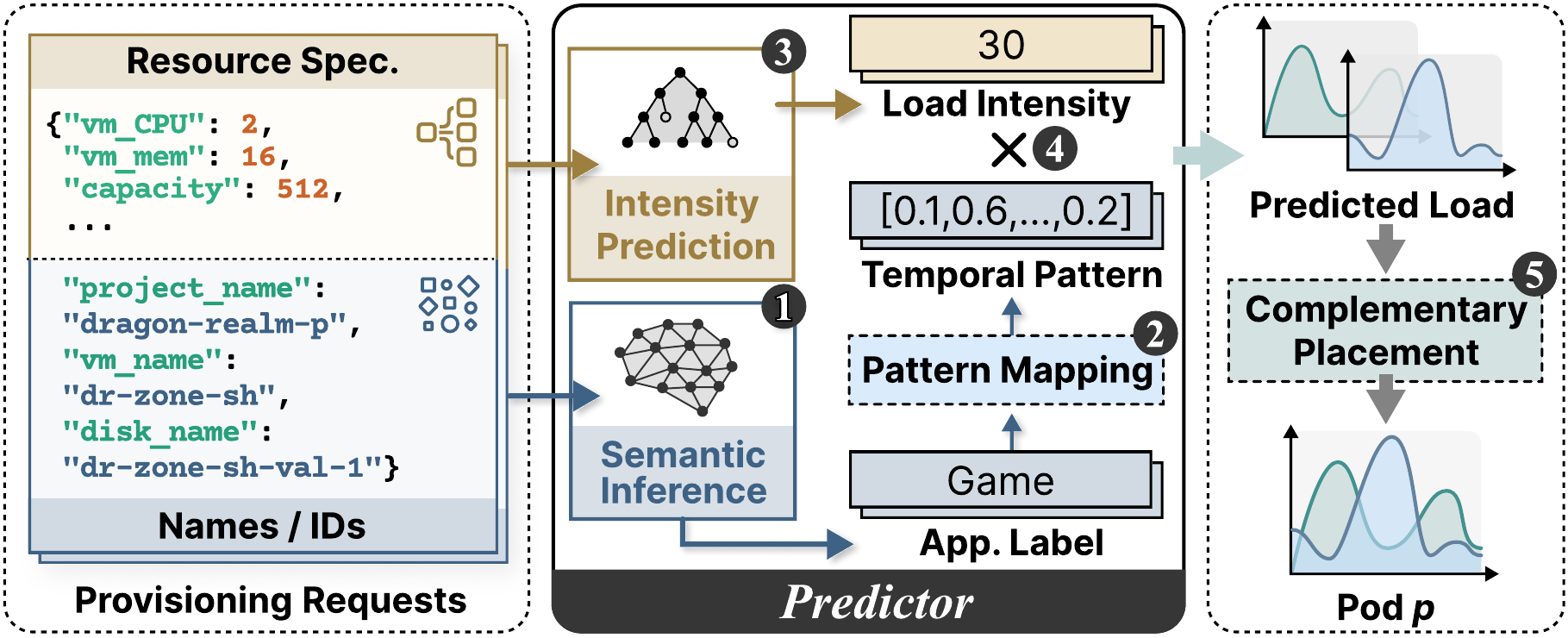}}
\caption{The end-to-end online workflow of TIDAL.}
\label{overview}
\end{figure}

Building on the analysis in §\ref{sec:bg} and §\ref{sec:opportunity}, we present \textbf{TIDAL}, a placement framework that exploits metadata-derived application semantics to enable complementary CVD placement. Figure~\ref{overview} illustrates its end-to-end workflow.

Upon receiving a provisioning request, TIDAL first infers an application label from tenant-provided identifiers in provisioning metadata (Step \ding{182}: Semantic inference). It then maps the label to a canonical temporal pattern (Step \ding{183}: Pattern mapping) and predicts load intensity from resource specifications (Step \ding{184}: Intensity prediction). These two components are combined into a predicted load profile for the new disk (Step \ding{185}: Profile synthesis), which the placement engine uses to select the pod whose aggregate load best complements that profile (Step \ding{186}: Complementary placement).

In TIDAL, Step~\ding{182} addresses \textbf{C1} by recovering application semantics from tenant-provided identifiers; we detail this stage in §\ref{sec:recovering}. Steps~\ding{183}--\ding{186} address \textbf{C2} by transforming the recovered semantics into a complementary placement decision; we present this process in §\ref{sec:semantic-to-placement}. To satisfy \textbf{C3}, we then introduce efficiency mechanisms in §\ref{sec:control-plane}; §\ref{sec:robustness} further describes safeguards for uncertainty in production deployment.

\subsection{Recovering Application Semantics}\label{sec:recovering}
To address \textbf{C1}, TIDAL first recovers application semantics from tenant-provided identifiers in provisioning metadata (Step~\ding{182}). Each request is represented as a tuple $M=\langle P,V,D\rangle$, where $P$, $V$, and $D$ denote the project, VM, and disk identifiers. The core difficulty is that these identifiers are noisy, open-vocabulary, and often non-literal. TIDAL therefore uses an offline teacher LLM to construct supervision and a lightweight online student model to perform stable semantic inference at runtime. Given $M$, this stage outputs an application label and confidence score $q_d$ for downstream placement logic.

\noindent\textbf{Data-driven taxonomy construction.}
Rather than fixing a label space manually, we build it from data. The teacher LLM first mines fine-grained application candidates from historical provisioning metadata, then consolidates categories that are both semantically adjacent and temporally indistinguishable---preserving only distinctions that matter for placement. The resulting taxonomy contains 27 representative classes (Table~\ref{tab:llm-coverage}).

\begin{figure}[t]
\centerline{\includegraphics[width=0.5\textwidth]{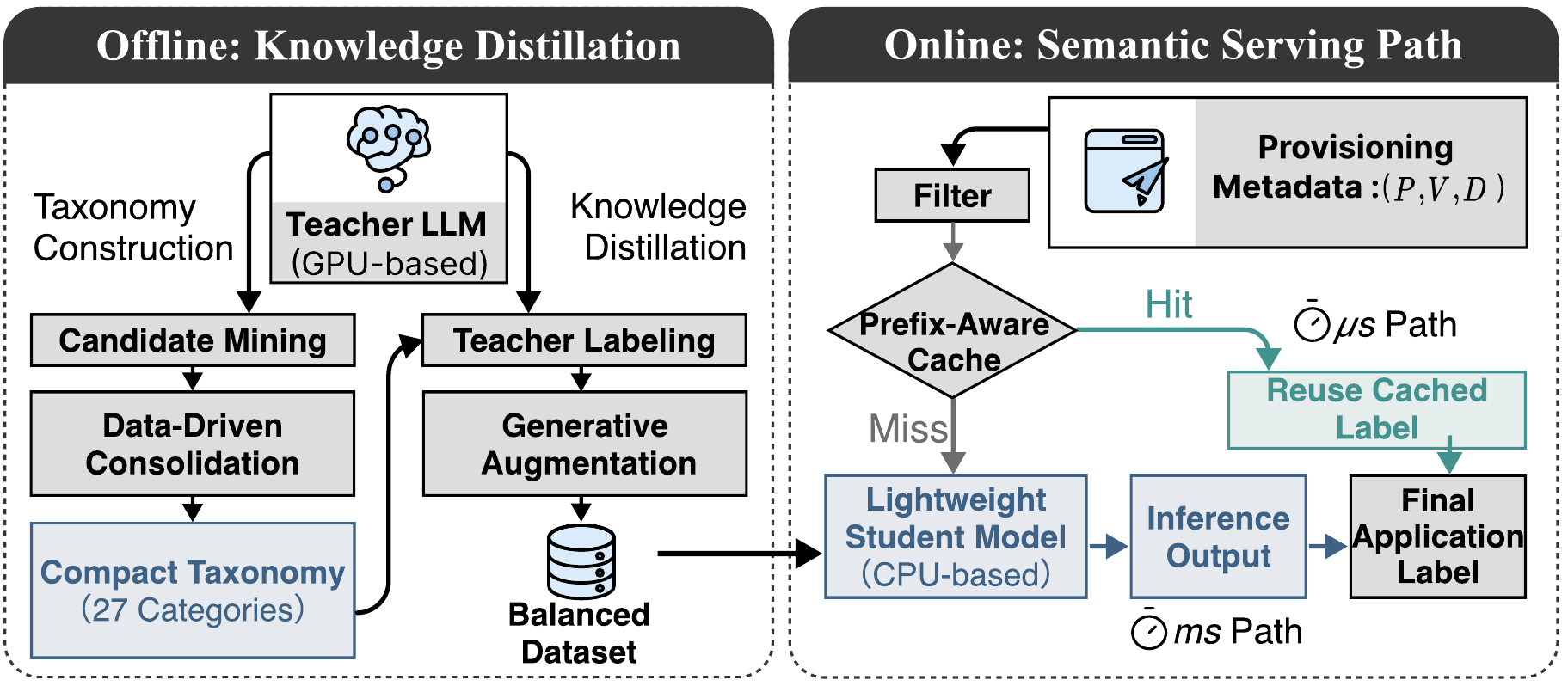}}
\caption{Offline-to-online semantic inference pipeline in TIDAL. The offline stage constructs the taxonomy and distills supervision from a teacher LLM, while the online stage serves semantic inference through filtering, prefix-aware caching, and a lightweight student model.}
\label{labeling}
\end{figure}

\noindent \textbf{Teacher-assisted supervision and augmentation.}
We next use the teacher LLM to annotate historical requests within this taxonomy, creating labeled supervision without placing the teacher on the provisioning critical path. The resulting data are highly imbalanced: a few generic classes dominate, while specialized applications form a long tail~\cite{disti_long_tail1}. To improve coverage on sparse classes, we further use the teacher to generate semantically equivalent variants of existing provisioning names, including aliases, abbreviations, and synonymous expressions. This augmentation improves class balance and strengthens the student model's robustness to naming diversity.

\noindent \textbf{Distilled online classifier.}
Downstream placement requires a stable semantic interface: each request must be mapped to one class in the fixed 27-class taxonomy together with a confidence score used by later control logic (\S~\ref{sec:robustness}). A prompted LLM does not naturally provide such an interface: its outputs may vary in wording, length, and format, may fall outside the taxonomy, and do not directly yield a normalized confidence score for fallback control. We therefore distill the teacher's semantic judgments into a lightweight supervised classifier whose output space is exactly the predefined taxonomy. At runtime, the student takes $M=\langle P,V,D\rangle$ as input and outputs $(\textit{label}, q_d)$. This converts noisy free-form identifiers into a compact, bounded, and machine-consumable semantic representation that can be translated into phase-aware signals in the next stage.

Figure~\ref{labeling} (left) illustrates this offline-to-online semantic pipeline. Once $(\text{label}, q_d)$ is obtained, TIDAL proceeds to infer the corresponding temporal signal for placement.

\subsection{From Semantics to Placement Decisions}
\label{sec:semantic-to-placement}
To address \textbf{C2}, TIDAL next transforms the recovered semantic label into a quantitative load profile and then into a concrete placement decision.

\begin{figure}[t]
\centerline{\includegraphics[width=0.47\textwidth]{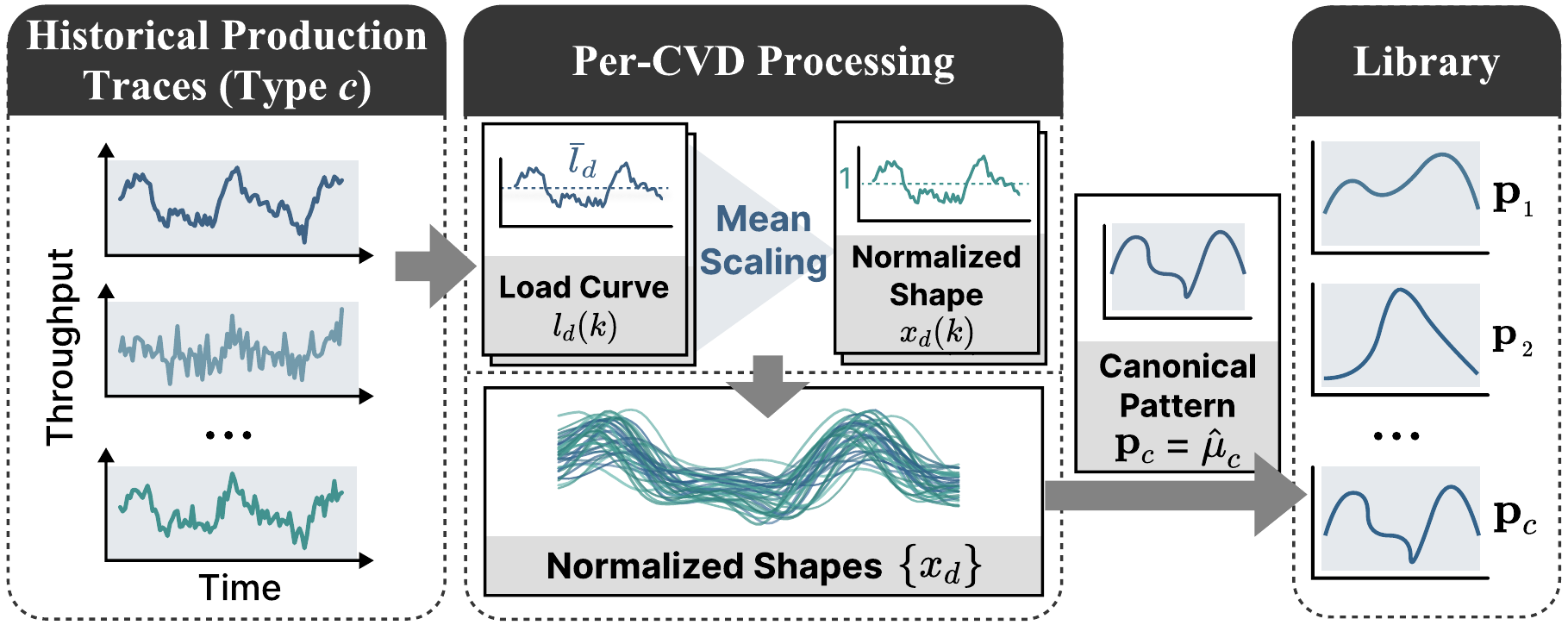}}
\caption{Construction process of the offline profile library.}
\label{design2}
\end{figure}

\noindent\textbf{Canonical temporal pattern (Step \ding{183}).} For each application class $c$, TIDAL constructs an offline profile library that stores a canonical intra-day temporal pattern $\mathbf{p}_c \in \mathbb{R}^{K}$. To build this library, we collect historical CVDs of class $c$, partition the 24-hour cycle into $K$ time slots, compute each disk's average load in each slot, and apply mean normalization to remove absolute scale. TIDAL then averages these normalized curves to obtain the canonical pattern $\mathbf{p}_c$ for class $c$. Figure~\ref{design2} illustrates this offline construction pipeline. At runtime, once a new CVD $d$ is inferred as class $c(d)$, TIDAL retrieves $\mathbf{p}_{c(d)}$ as its predicted temporal pattern. Note that our current design and evaluation focus on a single-region deployment setting, so the canonical pattern library is constructed and applied within one operational timezone.

This construction is justified by a concentration argument at pod scale: while individual disks of the same class may vary, their deviations tend to average out when many CVDs are aggregated, so the class-level aggregate load converges to a scaled version of the canonical pattern. We formalize this argument in Appendix~\ref{app:pattern-proof}.

\noindent\textbf{Load intensity prediction (Step \ding{184}).}
The semantic label captures \emph{when} load is likely to arrive, but not \emph{how much} load the CVD will generate. TIDAL therefore predicts a scalar load intensity $\hat{i}_d$ from resource specifications. Following prior work showing that load intensity correlates with these provisioning features~\cite{coach,VM_placement_rescen,TELA}, we use two categories of inputs: (1) CVM attributes, such as vCPU count and memory size; and (2) CVD attributes, such as provisioned capacity, lease duration, and disk role.

We formulate this task as regression rather than classification. Prior schemes often discretize intensity into coarse levels (e.g., Low/Medium/High) and formulate prediction as classification~\cite{power_pred,lava,SCDA,TELA}, but such quantization introduces packing error. Since overload avoidance depends on numerical mismatch against a fixed pod bandwidth budget, reducing numerical error is more important than predicting the correct bucket. Our evaluation confirms that regression better preserves this property and improves overload avoidance effectiveness (see §\ref{sec:split}).

\noindent\textbf{Profile synthesis (Step \ding{185}).}
TIDAL combines the predicted temporal shape and intensity into a full load profile for the new CVD:
\begin{equation}
\hat{\mathbf{l}}_d = \hat{i}_d \cdot \mathbf{p}_{c(d)}.
\label{eq:predicted-profile}
\end{equation}
This synthesized curve is the representation consumed by the placement engine. 

\noindent\textbf{Complementary placement (Step \ding{186}).}
For each pod $p$, the placement engine consults a current load vector $\mathbf{L}_p \in \mathbb{R}^K$ derived from recent backend monitoring data, which captures the pod's observed intra-day load shape. When a new CVD $d$ arrives, TIDAL evaluates the marginal effect of placing $\hat{\mathbf{l}}_d$ into each candidate pod. We quantify temporal volatility using intra-day variance:

\begin{equation}
\begin{split}
\mu(\mathbf{L}_p) = \frac{1}{K}\sum_{k=1}^{K} L_p(k), \\
\mathrm{var}(\mathbf{L}_p) = \frac{1}{K}\sum_{k=1}^{K}\big(L_p(k) - \mu(\mathbf{L}_p)\big)^2.
\end{split}
\end{equation}

The cost of placing $d$ into pod $p$ is:
\begin{equation}
\Delta \mathrm{var}_p(d) = \mathrm{var}(\mathbf{L}_p + \hat{\mathbf{l}}_d) - \mathrm{var}(\mathbf{L}_p).
\end{equation}

A negative value indicates that the CVD's peaks align with existing valleys and therefore improve temporal smoothing.

We model placement as an \emph{online vector bin-packing problem}~\cite{binpack1,binpack2}. Since CVD requests arrive as a continuous stream with unknown future arrivals, global combinatorial optimization is infeasible. TIDAL therefore adopts an $O(1)$ greedy heuristic policy (Algorithm~\ref{alg:tidal-placement}, Lines~5--12). To preserve basic spatial balance, it first forms a candidate set $\mathcal{P}_{\mathrm{cand}}$ consisting of the $M$ pods with the best post-placement spatial scores (see \S~\ref{sec:robustness}), and then chooses
\begin{equation}
p^* = \arg\min_{p \in \mathcal{P}_{\mathrm{cand}}} \Delta \mathrm{var}_p(d).
\label{eq:greedy}
\end{equation}

This policy preserves the efficiency required by the provisioning path while still exploiting temporal complementarity. As we show in \S~\ref{sec:meltdown}, the marginal-variance objective consistently outperforms alternative objectives such as minimizing post-placement variance or instantaneous peak load.

At the end of this stage, TIDAL has converted metadata-derived semantics into a concrete placement decision (Steps~\ding{183}--\ding{186}). The next two sections describe how this pipeline is made efficient enough for the control plane and robust to uncertainty in production deployment.


\begin{algorithm}[t]
\footnotesize
\caption{TIDAL Placement with Spatial Screening and Temporal Smoothing}
\label{alg:tidal-placement}
\DontPrintSemicolon

\KwIn{Disk size $s_d$, predicted intensity $\hat{i}_d$, canonical pattern $\mathbf{p}_c$, confidence $q_d$}
\KwData{Current per-pod spatial states $\mathcal{S}_p$; current per-pod temporal load vectors $\mathbf{L}_p$}
\KwParam{Confidence threshold $\tau$; candidate set size $M$}

\BlankLine

\If{$q_d < \tau$}{
    $p^\star \leftarrow \textsc{SpatialFallback}(s_d,\hat{i}_d,\{\mathcal{S}_p\})$\;
    \Return $p^\star$\;
}

$\hat{\mathbf{l}}_d \leftarrow \hat{i}_d \mathbf{p}_c$\;
$\mathcal{P}_{\mathrm{cand}} \leftarrow \textsc{SelectSpatialCandidates}(s_d,\hat{i}_d,\{\mathcal{S}_p\}, M)$\;
$\Delta^\star \leftarrow +\infty,\ p^\star \leftarrow \textsc{Nil}$\;

\ForEach{$p \in \mathcal{P}_{\mathrm{cand}}$}{
    $\Delta_p \leftarrow \mathrm{var}(\mathbf{L}_p + \hat{\mathbf{l}}_d) - \mathrm{var}(\mathbf{L}_p)$\;
    \If{$\Delta_p < \Delta^\star$}{
        $\Delta^\star \leftarrow \Delta_p$\;
        $p^\star \leftarrow p$\;
    }
}

\Return $p^\star$\;
\end{algorithm}
\subsection{Making TIDAL Practical}
\label{sec:control-plane}
To satisfy \textbf{C3}, TIDAL must make semantic recovery and its associated control logic lightweight enough to run on the provisioning critical path. TIDAL therefore incorporates several mechanisms to make the pipeline practical for deployment in the CBS control plane, as shown in Figure~\ref{labeling} (right).

\noindent\textbf{Distillation.}
The main step toward control-plane feasibility is to avoid direct LLM inference at provisioning time. As described in §\ref{sec:recovering}, TIDAL uses the teacher LLM only offline to construct supervision and train a lightweight student model. Besides producing stable semantic outputs, the distilled student removes the latency and infrastructure cost of online LLM serving, making semantic inference feasible on standard CPU-based control nodes.

\noindent\textbf{Regex-based filtering.}
Not all provisioning requests benefit from semantic inference. In production traces, a nontrivial fraction of metadata consists of deployment-generated strings that carry little recoverable semantic content, including hash/UUID-like identifiers, opaque alphanumeric tokens, and default system-assigned names (20.29\% in Table~\ref{tab:llm-coverage}). TIDAL therefore applies a lightweight regex-based filter before invoking the student model. Rather than attempting full semantic understanding, this filter only detects obvious non-semantic patterns, such as common system-default tokens, hash-like strings, and highly unnatural character compositions. Detailed rules and examples are provided in Appendix~\ref{app:noise_filter}. Requests intercepted by this filter bypass semantic inference and are routed directly to the degradation path in §\ref{sec:robustness}. This reduces unnecessary model invocations and prevents obviously meaningless strings from entering the semantic recovery stage.

\noindent\textbf{Prefix-aware semantic caching.}
Provisioning requests often exhibit strong locality. Tenants frequently create CVDs in batches under identical project and VM contexts, with disk names differing only by a systematic suffix such as an index or shard identifier (e.g., \verb|db-data-01|, \verb|db-data-02|). To exploit this reuse, TIDAL maintains a prefix-aware semantic cache over previously inferred requests. When a new request $M_{\text{new}}=\langle P_{\text{new}},V_{\text{new}},D_{\text{new}}\rangle$ arrives, TIDAL searches for a recent cached entry with the same project and VM identifiers and a sufficiently long longest common prefix (LCP) in the disk name. We declare a match when
\begin{equation}
\frac{\mathrm{len}(\mathrm{LCP}(D_{\text{new}},D_{\text{cached}}))}{\mathrm{len}(D_{\text{new}})} \ge \lambda.
\label{eq:lcp}
\end{equation}
Based on empirical sensitivity analysis, we set $\lambda=0.4$, which yields a 98.4\% probability that the two disks share the same application label. On a cache hit, TIDAL reuses the cached label and confidence score; otherwise it invokes the student model. The cache uses LRU eviction to exploit temporal locality while bounding memory usage.

\subsection{Robustness under Uncertainty}
\label{sec:robustness}
Beyond efficiency, production deployment must also account for uncertainty in ML-derived semantic signals. TIDAL therefore includes several safeguards to limit the impact of misclassification and stale offline knowledge.

\noindent\textbf{Confidence-based fallback.}
An incorrect semantic label may induce an incorrect temporal pattern and thus weaken complementary placement. To bound this risk, TIDAL uses the confidence score $q_d$ as a control signal. If $q_d < \tau$, TIDAL bypasses the phase-aware path and falls back to a multi-resource spatial balancing policy (Algorithm~\ref{alg:tidal-placement}, lines 1--3). This fallback uses the disk's provisioned capacity and predicted load intensity to balance capacity and load utilization across pods (similar to methods in \cite{SCDA}), while discarding only the unreliable phase-aware signal. We apply the same fallback to requests intercepted by the regex filter in §\ref{sec:control-plane}. The same spatial criterion is also used to select candidate pods for phase-aware placement, while the temporal objective is applied only within that candidate set. 
Detailed spatial scoring rules for both candidate selection and fallback are deferred to Appendix~\ref{app:spatial_fallback}.

\noindent\textbf{Monitoring-grounded pod state.}
TIDAL does not maintain pod load vectors by recursively accumulating predicted profiles of past placements. Instead, before each placement decision, the scheduler consults the pod's current load vector $\mathbf{L}_p$, derived from recent backend monitoring data and reflecting the pod's actual observed load. This design helps contain prediction errors: even if an earlier request was mislabeled or suboptimally placed, its real load contribution will appear in subsequent observations, so later decisions are made against measured pod state rather than stale predicted accumulation. As a result, prediction errors do not recursively distort the scheduler's internal view of cluster load.

\noindent\textbf{Profile-library refresh.}
The canonical pattern library is constructed offline from historical traces and should be refreshed periodically to track long-term shifts in workload behavior. In deployment, this can be implemented using a rolling trace window over recent telemetry. These refreshes update only the offline library and do not interfere with the online placement path.

\subsection{Implementation}\label{sec:implement}
We implement TIDAL as a prototype CBS placement framework in 2,400 lines of Python. Key implementation details are as follows.

\noindent \textbf{Hyperparameters.} We set the temporal resolution to $K=12$ (2-hour time slots). For complementary placement (§\ref{sec:semantic-to-placement}), the spatial pre-filter considers $M=4$ candidate pods. The confidence threshold for degradation (§\ref{sec:robustness}) is set to $\tau=0.6$.

\noindent \textbf{Model architecture.} (i) Semantic distillation. We use a locally deployed DeepSeek-R1 \cite{llm_deepseek} as the offline teacher and DistilBERT \cite{distillbert} as the online student model. The prefix-aware cache is implemented as a hash map with an LRU eviction policy and a capacity of 10,000 entries ($\approx$20\% testset size). (ii) Intensity prediction. We employ a RandomForest regressor \cite{randomfor} via scikit-learn \cite{scikit-learn}. 

\noindent \textbf{Training data.} Our models are trained on 14,000 CVDs from 16 clusters of CloudProvider-A, containing both provisioning metadata and historical I/O traces. This training set is strictly disjoint from the testing data used in §\ref{sec:evaluation}.

\begin{figure*}[t]
\centerline{\includegraphics[width=1\textwidth]{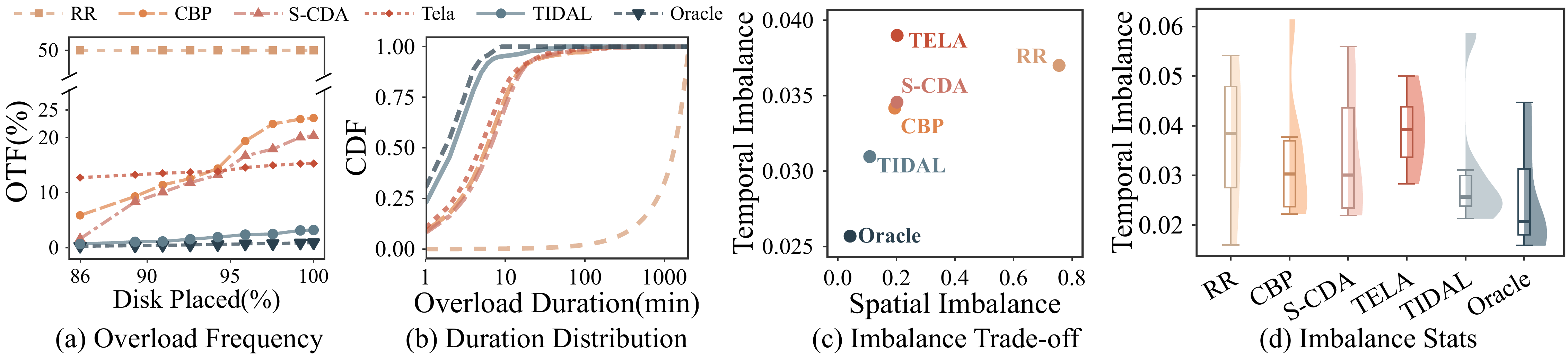}}
\caption{Effectiveness of TIDAL in load smoothing. (a) Overload time fraction (OTF) as placement progresses. (b) CDF of overload duration. (c) Spatial vs. temporal imbalance. (d) Distribution of temporal imbalance across pods.}
\label{eva1}
\end{figure*}

\section{Evaluation}\label{sec:evaluation}
\subsection{Experimental Setup}
\label{sec:setup}

We evaluate TIDAL along two dimensions: (1) \emph{effectiveness}, using trace-driven simulation to quantify load smoothing and overload mitigation; and (2) \emph{deployability}, by measuring runtime overhead on a standard CPU-only server.

\noindent \textbf{Trace-driven simulator.} We build a discrete-event simulator to evaluate placement policies at production scale without perturbing the live system. 
The workload is derived from production traces of 52{,}000 real CVDs collected over a two-week period in May 2023 from CloudProvider-A. Our study focuses on a single-region deployment setting, so both profile construction and evaluation are performed within one operational timezone rather than across mixed regions. This evaluation set is strictly disjoint from the datasets used for offline profile construction and model training (§\ref{sec:implement}), eliminating data leakage.

The simulator replays provisioning requests in chronological order. Once a CVD is assigned to a pod, the simulator replays its historical I/O trace from the offset matching the simulated time-of-day, rather than from the trace beginning, and aggregates throughput every 5 minutes to update the pod-level load vector $\mathbf{L}_p$. This preserves the original phase relationship among co-running workloads and captures whether their peaks overlap or interleave after placement. We model a cluster with 16 storage pods. To emulate a moderately constrained production environment, we set the aggregate bandwidth budget to $1.2\times$ the average throughput of the entire workload. An overload event is defined as any 5-minute interval in which a pod's aggregate throughput exceeds its bandwidth capacity.

\noindent \textbf{Hardware configuration.}
All experiments are conducted on a dedicated server with 128 AMD CPU cores (2.9\,GHz), 256\,GB DRAM, and a 2\,TB NVMe SSD. For accurate latency measurement, we isolate the TIDAL placement engine and the trace-replay process using \verb|cgroups| and CPU affinity, and pin them to separate physical cores on different NUMA nodes.

\noindent \textbf{Baselines.}
We compare TIDAL against five baselines:

\bulletpara{Round-Robin (RR):} Assigns CVDs to pods in strict rotation, serving as the simplest spatial distribution baseline.
\bulletpara{Capacity-Based Placement (CBP):} Assigns the CVD to the pod with the lowest current storage capacity utilization.
\bulletpara{S-CDA}\cite{SCDA}: Predicts each CVD's average load intensity from its configuration and balances average load across pods.
\bulletpara{TELA}\cite{TELA}: Identifies bursty CVDs and scatters them to prevent peak collisions, but notably lacks awareness of when bursts occur.
\bulletpara{Oracle:} An upper bound that assumes perfect knowledge of each arriving CVD's future temporal load, but still evaluates placement sequentially in the same request order and under the same greedy policy as TIDAL.

\subsection{Effectiveness in Load Smoothing}
\label{sec:effect}
We first evaluate whether TIDAL reduces overloads at the cluster level using two metrics: (1) \textbf{Overload Time Fraction (OTF)}, the fraction of runtime during which a pod's aggregate throughput exceeds its capacity; and (2) \textbf{Overload Duration}, the length of each continuous overload episode.

\noindent \textbf{Reduction in overload frequency.} 
Figure~\ref{eva1}(a) plots OTF as the placed fraction of CVDs increases from 86\% to 100\%. CBP and S-CDA also deteriorate as utilization rises, reaching around 20\% OTF at full placement. This confirms that balancing capacity or average load alone is insufficient once temporally aligned peaks begin to dominate aggregate load. TELA performs better by dispersing bursty disks, but TIDAL consistently achieves the lowest OTF among all practical schemes, with only 3.19\% OTF at full placement, and closely tracking Oracle. At 100\% placement, TIDAL reduces OTF by 79.1\% relative to TELA.

\noindent \textbf{Mitigation of long-tail overloads.} 
Figure~\ref{eva1}(b) shows the CDF of overload duration on a log-scaled x-axis. RR exhibits a pronounced heavy tail, indicating frequent long-lasting congestion. TELA shortens these episodes, but TIDAL shifts the entire distribution further left: most overloads become short-lived, while persistent overloads are largely eliminated. Relative to TELA, TIDAL reduces P95 and P99 overload duration by 73.7\% and 62.6\%, respectively. The remaining gap from TIDAL to Oracle is much smaller than the gap from TIDAL to any non-semantic baseline, suggesting that phase awareness is the main missing signal in prior approaches.

\noindent \textbf{Why spatial balance alone is insufficient.} To explain these gains, we analyze the final cluster state along two dimensions: Spatial Imbalance, measured as the CoV of average load across pods, and Temporal Imbalance, measured as the average intra-pod CoV of load over time. Figure~\ref{eva1}(c) places each policy in this trade-off space. TELA lies in the upper-left region: it achieves low spatial imbalance, but still leaves substantial temporal imbalance. Scattering bursty disks helps avoid some direct peak collisions, yet without phase awareness it cannot actively interleave peaks with valleys, so pod-level load curves remain volatile.

TIDAL moves the operating point down and left simultaneously. It reduces temporal imbalance by recovering phase information from semantic labels, while also achieving slightly better spatial balance than S-CDA. This secondary gain comes from prediction granularity: S-CDA relies on coarse load categories, whereas TIDAL uses scalar intensity estimates (§\ref{sec:semantic-to-placement}), enabling more accurate balancing of average traffic mass during greedy placement.

Figure~\ref{eva1}(d) further supports this interpretation. TIDAL achieves the lowest median temporal imbalance among all practical schemes and exhibits a tighter spread across pods, second only to Oracle. In other words, TIDAL not only keeps load under the threshold more often, but also produces intrinsically smoother pod-level load curves.

\subsection{Dissecting TIDAL's Gains}
\label{sec:meltdown}
We next isolate which components are responsible for TIDAL's gains.

\begin{figure}[t]
\centerline{\includegraphics[width=0.47\textwidth]{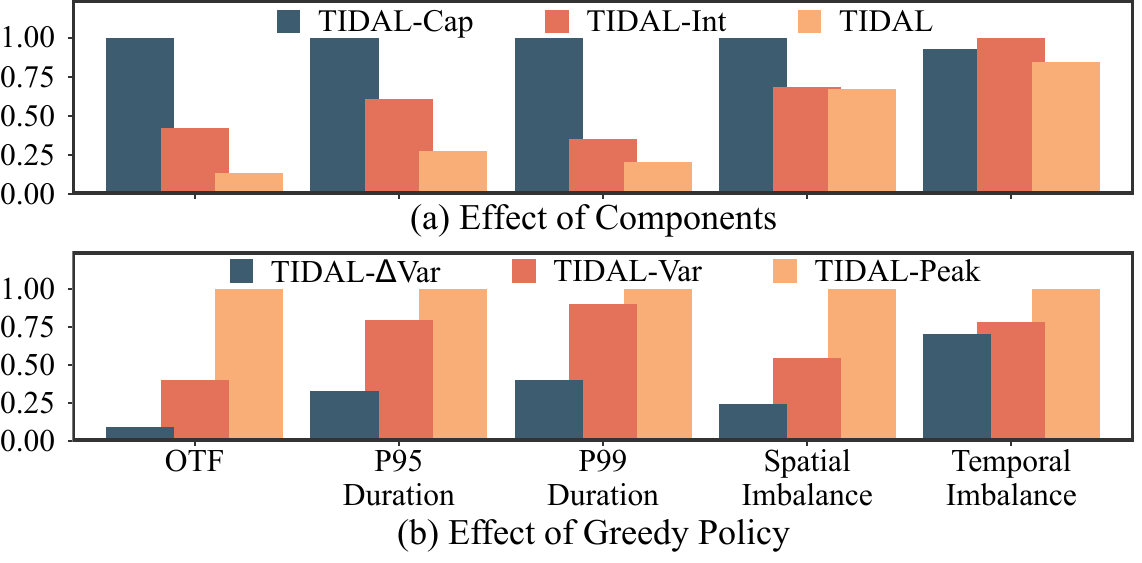}}
\caption{Ablation study. (a) Impact of intensity prediction and semantic inference. (b) Comparison of greedy objectives.}
\label{eva2}
\end{figure}

\noindent \textbf{Impact of system components.} 
We compare three progressively stronger variants: (i) \textbf{TIDAL-Cap}, which uses only capacity-based placement; (ii) \textbf{TIDAL-Int}, which adds the intensity regressor but not semantic labels; and (iii) \textbf{TIDAL}, the full design with both intensity prediction and semantic inference. Figure~\ref{eva2}(a) shows that these two components contribute in different ways. Moving from TIDAL-Cap to TIDAL-Int reduces Spatial Imbalance by 31\%, confirming that scalar intensity prediction improves average-load packing beyond simple capacity counts. The decisive gain appears when moving from TIDAL-Int to full TIDAL: \textbf{adding semantic signals reduces OTF by more than $3\times$.} Intensity prediction improves \emph{how much} load is placed on each pod, while semantic inference determines \emph{when} that load is likely to arrive, enabling phase-complementary placement that suppresses aggregate peaks.

\noindent \textbf{Sensitivity to greedy objective.} 
We then evaluate the objective used by the greedy placement engine. We compare three variants: (i) \textbf{TIDAL-$\Delta$Var}, which minimizes the incremental change in intra-day variance (default); (ii) \textbf{TIDAL-Var}, which minimizes the absolute variance after placement; and (iii) \textbf{TIDAL-Peak}, which minimizes peak load.

Figure~\ref{eva2}(b) shows that TIDAL-$\Delta$Var performs best overall. TIDAL-Peak performs worst, indicating that myopically avoiding the current maximum does not produce smooth load shapes over time. TIDAL-Var performs better, but still trails TIDAL-$\Delta$Var. The reason is that $\Delta$Var directly rewards temporal complementarity: it prefers placements whose predicted peaks fill existing valleys, rather than merely reducing a snapshot statistic after placement. This objective is therefore best aligned with TIDAL's goal of smoothing pod-level load curves and preventing overloads.


\subsection{Accuracy of ML Components} 
\label{sec:split}
We evaluate the two learned components in isolation to justify the model choices in TIDAL.

\noindent \textbf{Semantic recovery.}
We compare nine Transformer-based student models distilled from the teacher LLM, including BERT-base \cite{bert-base}, DistilBERT \cite{distillbert}, RoBERTa-base \cite{xlm-roberta}, RBT3 / RBTL3 from BERT-wwm \cite{bert-wwm}, MiniRBT \cite{miniRBT}, MobileBERT \cite{mobilebert}, CNMBERT \cite{cnmbert}, and TinyBERT \cite{tinybert}. We also include two non-Transformer baselines: FastText \cite{fasttext} and a TF-IDF + logistic-regression classifier \cite{scikit-learn} augmented with manually curated keyword lists (10--12 keywords per type).

\begin{figure}[t]
\centerline{\includegraphics[width=0.47\textwidth]{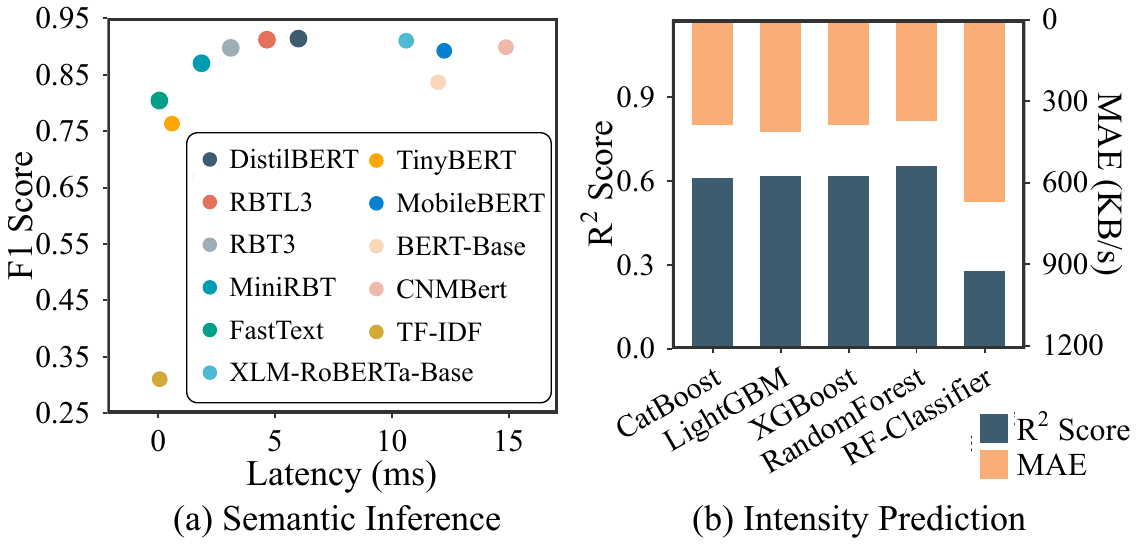}}
\caption{Accuracy of ML components. (a) F1 vs. CPU latency for semantic inference models. (b) $R^2$ and MAE for intensity prediction models.}
\label{eva3}
\end{figure}

Figure~\ref{eva3}(a) plots F1 score against CPU inference latency. The lexical baseline performs poorly: TF-IDF reaches 0.34 F1, indicating that simple keyword matching is insufficient for recovering semantics from noisy metadata. In contrast, even FastText trained on teacher-labeled data reaches 0.8 F1 with millisecond-scale latency. Transformer-based student models dominate the Pareto frontier, showing that the teacher's semantic supervision can be distilled into lightweight online classifiers. Among them, DistilBERT achieves the best overall trade-off, reaching 0.92 F1 while keeping inference latency in the single-digit millisecond range. We therefore use DistilBERT as the default student model.

\noindent \textbf{Intensity prediction.} 
We next compare four tree-based regressors—CatBoost\cite{catboost}, LightGBM\cite{lightgbm}, XGBoost\cite{xgboost}, and RandomForest\cite{randomfor}—against a classification baseline that first discretizes intensity into four coarse buckets and then applies a RandomForest classifier. Figure~\ref{eva3}(b) reports the coefficient of determination ($R^2$) and mean absolute error (MAE).

All four regressors achieve comparable accuracy, with $R^2$ around 0.6--0.65 and MAE in the 400--450\,KB/s range, indicating that resource specifications are informative for average-intensity estimation. RandomForest slightly outperforms the others on both metrics and therefore provides the best overall fit. In contrast, the classification baseline performs markedly worse: despite achieving 0.84 F1 on the bucket labels, its $R^2$ is less than half that of the regressors and its MAE nearly doubles. This confirms that coarse discretization introduces substantial quantization error even when bucket-level classification appears accurate, and explains why scalar regression enables more accurate spatial balancing than S-CDA-style categorization (§\ref{sec:effect}).

\subsection{Deployability and Robustness}
\label{sec:robust}
\noindent \textbf{Online latency.}
Table~\ref{tab:latency} reports the end-to-end placement latency on our CPU-only server. Simple spatial policies incur only $3.4$--$3.7\,\mathrm{ms}$ per CVD request. Without online optimizations, TIDAL takes $11.40\,\mathrm{ms}$, with the extra cost dominated by semantic inference. Prefix-aware caching reduces this overhead to $8.63\,\mathrm{ms}$ (TIDAL-c), a 24.3\% reduction. Adding regex-based filtering further lowers latency to $7.03\,\mathrm{ms}$ (TIDAL-cf), reducing the uncached cost by 38.3\% overall. This remains well within the tens-of-milliseconds control-plane budget reported in industrial practice~\cite{vm_kamino}. Combined with the gains in Figure~\ref{eva1}, these results indicate a favorable control-plane trade-off: a few milliseconds of extra provisioning cost substantially reduce runtime overload.

\begin{figure}[t]
\centerline{\includegraphics[width=0.48\textwidth]{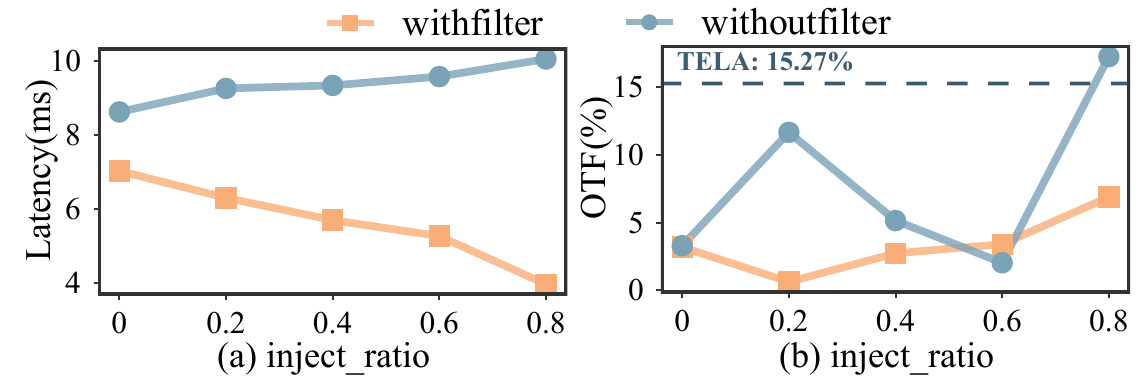}}
\caption{Robustness to semantically weak metadata. (a) Placement latency and (b) OTF under injected meaningless metadata, with and without regex-based filtering.}
\label{eva4}
\end{figure}

\begin{table}[t]
  \centering
  \caption{Online placement latency per provisioning request.}
  \label{tab:latency}
  \scriptsize                 
  \setlength{\tabcolsep}{2pt} 
  \renewcommand{\arraystretch}{0.9}%
  \resizebox{\columnwidth}{!}{
    \begin{tabular}{lccccccc}
      \toprule
      Scheme & TIDAL & TIDAL-c & TIDAL-cf & RR & S\mbox{-}CDA & TELA & Oracle \\
      \midrule
      \emph{ms} / req
             & 11.40 & 8.63    & 7.03 & 3.41 & 3.38       & 3.66 & 9.75 \\
      \bottomrule
    \end{tabular}
  }
\end{table}

\noindent \textbf{Robustness to semantically weak metadata.}
We next stress-test TIDAL by injecting random meaningless strings into provisioning metadata, simulating requests whose names carry little recoverable semantics. Figure~\ref{eva4}(a) shows the resulting latency. Without filtering, latency increases because randomized names reduce prefix-cache reuse and trigger more ineffective semantic inference. In contrast, with regex-based filtering enabled, latency decreases steadily as more injected requests are intercepted before student inference.

Figure~\ref{eva4}(b) shows the corresponding placement quality. Even without filtering, TIDAL remains reasonably robust because confidence-based fallback routes 81--82\% of randomized requests to the intensity-only path, preserving intensity-aware spatial balancing. However, without filtering, OTF becomes highly variable and eventually exceeds TELA under heavy noise. With filtering enabled, TIDAL remains much more stable across all injection ratios and consistently outperforms TELA; regex-based filtering intercepts 92--93\% of randomized requests. This shows that regex-based filtering is not merely a latency optimization: it also protects the semantic recovery pipeline from obvious non-semantic metadata and stabilizes degradation from intensity+phase placement to intensity-only placement.

\begin{figure}[t]
\centerline{\includegraphics[width=0.48\textwidth]{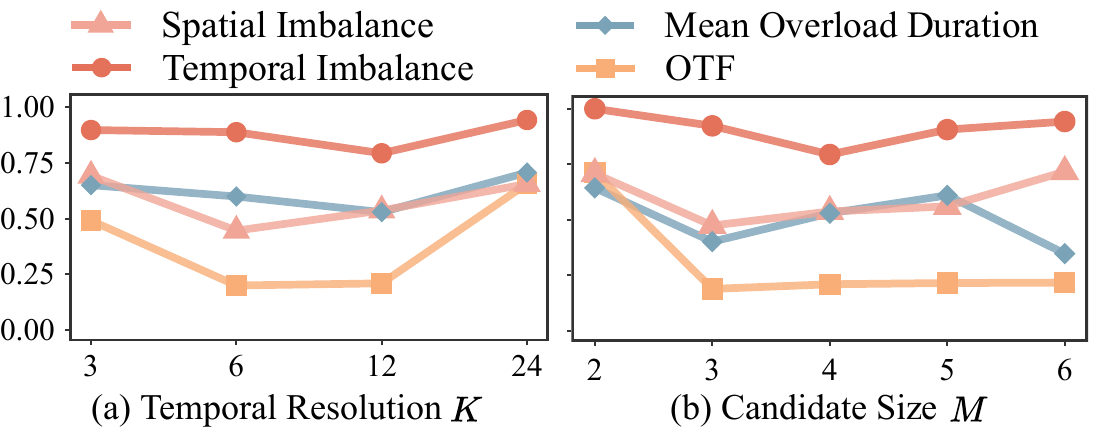}}
\caption{Hyperparameter sensitivity. (a) Sensitivity to temporal resolution $K$. (b) Sensitivity to candidate size $M$.}
\label{eva5}
\end{figure}

\noindent \textbf{Hyperparameter sensitivity.}
We finally examine sensitivity under full placement by varying two key parameters. All metrics in Figure~\ref{eva5} are normalized to the best baseline TELA. For temporal resolution $K$, Figure~\ref{eva5}(a) shows a clear U-shaped trend: small $K$ is too coarse to distinguish neighboring phases, whereas large $K$ overfits fine-grained noise. Intermediate settings ($K=6$ or $12$) provide the best balance. For candidate set size $M$, Figure~\ref{eva5}(b) shows that a modest candidate pool captures most of the benefit. Increasing $M$ from 2 to 3 sharply reduces overload, while larger values keep OTF low but gradually increase spatial imbalance as the algorithm gains more freedom to prioritize temporal valleys over average-load balance. Overall, TIDAL is robust across a broad parameter range, and our default settings lie near the best operating region.

\section{Related Work}
\label{sec:related}

\noindent\textbf{VM management.}
A large body of work studies VM management, spanning VM placement~\cite{vm_protean,vm_pMapper,vm_kamino,coach,vm_cha,lava,VM_placement_TAPAS}, migration~\cite{vm_migration,vm_blackbox,vm_livemig}, and consolidation~\cite{vm_entropy,vm_harvest,vm_aggregate,vm_smarthavest}. These systems primarily aim to improve resource utilization and load balance, thereby enhancing both provider efficiency and service quality.

\medskip
\noindent\textbf{Cloud load balancing.}
\emph{Front-end balancing:} Systems such as~\cite{balance_ananta,balance_maglev,balance_silkroad,balance_beamer,balance_bar} focus on connection steering and service fan-out. They use mechanisms such as consistent hashing and ASIC offload to maximize throughput while maintaining per-connection consistency. \emph{Storage-backend balancing:} Systems such as~\cite{balance_Kurma,EBS_DPU,SCDA,TELA,EBS_evolut} focus on balancing persistent storage traffic and data placement. These schemes couple load balancing with data layout or I/O scheduling policies.

\medskip
\noindent\textbf{ML for Systems.}
ML-driven resource management is often built as predict-then-act~\cite{Martin_survey}: a model predicts a system signal and a decision module consumes that prediction~\cite{coach,microserv_sinan,VM_placement_rescen,ML_job_rubick,SCDA,TELA,LLAMA,ML_job_Optimus,ML_job_Pollux}. Other systems learn the policy directly~\cite{George,olpart,fleetio,Utilize_protorail,microserv_firm} or replace the control loop with reinforcement learning~\cite{rl_manage,rl_sche,rl_ms}. TIDAL follows the predict-then-act framework, but extends the observation space by incorporating unstructured tenant-provided identifiers in provisioning metadata to recover application semantics and derive phase-aware placement signals.
\section{Discussion}
\label{sec:discussion}

\noindent\textbf{Privacy and metadata usage.}
TIDAL operates only on provisioning metadata in the control plane and never inspects tenant data content. To further reduce privacy risk, our training and ingestion pipeline was preceded by lightweight sanitization~\cite{anonymization}.

\noindent\textbf{Workload drift and migration.}
Like any initial placement scheme, TIDAL cannot eliminate workload drift: a CVD's temporal behavior may later diverge from its initial semantics. In practice, long-term balance still requires online monitoring and occasional migration. TIDAL improves the starting point by producing better initial placements, thereby reducing the need for subsequent migration when migration is expensive or infrequent.

\noindent\textbf{Beyond cloud block storage.}
Although we focus on CBS, the core idea is broader: control-plane identifiers often encode workload semantics. Similar signals appear in other systems, such as \texttt{job\_name} fields in cluster schedulers like Kubernetes~\cite{Kubernetes} and Slurm~\cite{SLURM}, or function identifiers in serverless platforms. Extending TIDAL's semantic-to-signal pipeline to these domains is a direction for future work.

\noindent\textbf{Limitations and future work.}
TIDAL has three main scope boundaries. First, our current study focuses on a single-region deployment setting, so both profile construction and online placement operate within one operational timezone. Extending TIDAL to multi-region deployments would require explicit timezone-aware handling. Second, the class-level canonical pattern is most reliable when load is distributed across many disks within a semantic class; it becomes less accurate when a few extremely high-intensity disks dominate the load mass. Third, TIDAL currently uses one canonical pattern per semantic class, whereas real workloads may contain multiple subtypes with distinct temporal behavior. Exploring finer semantic subtyping or multi-pattern representations is a natural direction for future work.
\section{Conclusion}
We presented TIDAL, a CVD placement framework that exploits application semantics encoded in provisioning metadata to enable complementary placement. By combining LLM-guided semantic recovery, an offline profile library, and lightweight intensity prediction, TIDAL infers a phase-aware load profile for each new disk at provisioning time and places it to smooth pod-level load. Trace-driven experiments on production workloads show that TIDAL substantially reduces overload risk and temporal imbalance compared to spatial baselines, while maintaining practical control-plane latency.

\begin{acks}
We thank Tencent Cloud for generously providing large-scale production data that enabled this study. Their support was essential to our empirical analysis and system evaluation. 
\end{acks}




\bibliographystyle{ACM-Reference-Format}
\bibliography{sample-base}

\appendix

\section{Statistical Justification for Canonical Pattern Aggregation}
\label{app:pattern-proof}

This appendix justifies the canonical-pattern construction used in \S\ref{sec:semantic-to-placement}.

For a given application class $c$, let $\mathcal{D}_c$ denote the set of historical CVDs belonging to this class. We partition the 24-hour cycle into $K$ time slots. For each disk $d \in \mathcal{D}_c$, we compute its average load in slot $k$, denoted $l_d(k)$, yielding a raw intra-day load curve $\mathbf{l}_d = [l_d(1), \dots, l_d(K)]$.

To separate temporal \emph{shape} from absolute scale, we apply mean normalization:
\begin{equation}
\begin{split}
\mathbf{x}_{d} = \left[x_{d}(1), \ldots, x_{d}(K)\right], \\
\quad \text{where } x_{d}(k) = \frac{l_{d}(k)}{\bar{l}_{d}}, \quad \bar{l}_{d} = \frac{1}{K} \sum_{k=1}^{K} l_{d}(k)
\end{split}
\end{equation}

We treat $\mathbf{x}_d$ as a realization of a random vector $\mathbf{X}_c$ associated with class $c$. The canonical temporal pattern of class $c$ is approximated by the sample mean
\begin{equation}\label{eq2}
\hat{\boldsymbol{\mu}}_c = \frac{1}{|\mathcal{D}_c|} \sum_{d \in \mathcal{D}_c} \mathbf{x}_d,
\end{equation}
which estimates the population mean $\boldsymbol{\mu}_c = \mathbb{E}[\mathbf{X}_c]$. In~\S\ref{sec:semantic-to-placement}, we denote this canonical pattern by $\mathbf{p}_c$.

In~\S\ref{sec:semantic-to-placement},  this $\hat{\mu}_c$ is designated as the \emph{canonical temporal pattern} $\mathbf{p}_c$. To justify this approximation, model the normalized shape of an individual CVD $d$ in class $c$ as:

\begin{equation}
\mathbf{x}_d = \mathbf{p}_c + \boldsymbol{\varepsilon}_d,
\quad \text{s.t.} \quad \mathbb{E}[\boldsymbol{\varepsilon}_d] = \mathbf{0}
\end{equation}

where $\boldsymbol{\varepsilon}_d$ captures disk-level deviation around the class pattern.

Consider a pod hosting $N$ class-$c$ CVDs. Its aggregate contribution from this class is
\begin{equation}
\mathbf{L}_{N,c} = \sum_{d=1}^{N} \mathbf{l}_d = \sum_{d=1}^{N} \bar{l}_d\,\mathbf{x}_d = \underbrace{\left(\sum_{d=1}^{N} \bar{l}_d\right)}_{\text{Signal Magnitude}} \mathbf{p}_c + \underbrace{\sum_{d=1}^{N} \bar{l}_d \boldsymbol{\varepsilon}_d}_{\text{Noise Term}}
\end{equation}

The aggregate load thus decomposes into a \emph{signal} term and a \emph{noise} term. Since the noise is a weighted sum of zero-mean random vectors, its magnitude grows sub-linearly at $O(\sqrt{N})$, whereas the signal term grows linearly $O(N)$. Consequently, as the number of hosted CVDs increases, the relative effect of individual deviations diminishes, and the aggregate load converges toward a scaled version of the canonical pattern:
\begin{equation}\label{eq5}
\mathbf{L}_{N,c} \approx 
\underbrace{\left( \sum_{d=1}^{N} \bar{l}_d \right)}_{I_c} \cdot \mathbf{p}_c
\end{equation}
where $I_c$ is the aggregate intensity of class $c$ in the pod.

This is the property exploited by TIDAL. The design does not require all disks in a class to share identical traces. It only requires class-level aggregate behavior to be sufficiently stable at pod scale, where many CVDs are statistically superposed. In this regime, the canonical pattern provides a robust approximation for complementary placement.

\section{Spatial Candidate Selection and Fallback Scoring}
\label{app:spatial_fallback}

TIDAL uses the same spatial balancing principle in two places: (1) to select the candidate pods for phase-aware placement, and (2) to place requests whose semantic signal is unavailable or unreliable. In both cases, the objective combines two terms: \emph{multi-resource balance within each pod} and \emph{spatial balance across pods}.

\noindent\textbf{Per-pod post-placement utilization.}
For each pod $p$, let $C_p$ and $L_p$ denote its current used capacity and observed average load, and let $C_p^{\max}$ and $L_p^{\max}$ denote its capacity and bandwidth limits, respectively. For an incoming disk $d$, its requested size $s_d$ is known from the provisioning request, and its predicted load intensity is $\hat{i}_d$.

If disk $d$ is tentatively placed into pod $p$, the resulting utilization ratios are
\begin{equation}
\tilde{u}_j^{\mathrm{cap}}(p)=
\begin{cases}
\dfrac{C_j+s_d}{C_j^{\max}}, & j=p,\\[6pt]
\dfrac{C_j}{C_j^{\max}}, & j\neq p,
\end{cases}
\qquad
\tilde{u}_j^{\mathrm{load}}(p)=
\begin{cases}
\dfrac{L_j+\hat{i}_d}{L_j^{\max}}, & j=p,\\[6pt]
\dfrac{L_j}{L_j^{\max}}, & j\neq p.
\end{cases}
\label{eq:post_utilization}
\end{equation}

\noindent\textbf{(1) Multi-resource balance within a pod.}
A pod is well balanced when its capacity utilization and load utilization are close. Intuitively, a pod with $(0.8, 0.8)$ is better balanced than one with $(0.95, 0.35)$, because the latter strands one resource while stressing the other. For candidate pod $p$, we therefore define its intra-pod imbalance as
\begin{equation}
B_{\mathrm{intra}}(p)=
\left|
\tilde{u}_p^{\mathrm{cap}}(p)-\tilde{u}_p^{\mathrm{load}}(p)
\right|.
\label{eq:intra_balance}
\end{equation}
Smaller values indicate better multi-resource balance and lower risk of resource stranding.

\noindent\textbf{(2) Spatial balance across pods.}
Besides balancing the two resource dimensions within one pod, placement should also avoid concentrating usage on a few hot pods. We measure this using the CoV of post-placement utilization across all pods:
\begin{equation}
B_{\mathrm{inter}}(p)=
\mathrm{CoV}\!\left(\{\tilde{u}_j^{\mathrm{cap}}(p)\}_{j=1}^{N}\right)
+
\mathrm{CoV}\!\left(\{\tilde{u}_j^{\mathrm{load}}(p)\}_{j=1}^{N}\right),
\label{eq:inter_balance}
\end{equation}
where $N$ is the number of pods. Smaller values indicate that capacity usage and load usage are more evenly distributed cluster-wide.

\noindent\textbf{Combined spatial score.}
We combine the above two terms into a single post-placement spatial score:
\begin{equation}
\mathrm{Score}_{\mathrm{spatial}}(p)
=
\lambda\, B_{\mathrm{intra}}(p)
+
(1-\lambda)\, B_{\mathrm{inter}}(p),
\label{eq:spatial_score}
\end{equation}
where $\lambda\in[0,1]$ controls the trade-off between intra-pod multi-resource balance and cluster-wide spatial balance. 

\noindent\textbf{Candidate selection.}
In the normal phase-aware path, TIDAL first ranks all pods by $\mathrm{Score}_{\mathrm{spatial}}(p)$ and selects the top-$M$ pods with the lowest scores:
\begin{equation}
\mathcal{P}_{\mathrm{cand}}
=
\operatorname*{arg\,topM}_{p}
\left(-\mathrm{Score}_{\mathrm{spatial}}(p)\right).
\label{eq:topm_spatial}
\end{equation}
Temporal smoothing is then applied only within this candidate set.

\noindent\textbf{Fallback placement.}
When semantic confidence is below threshold, or when a request is intercepted by the regex filter, TIDAL bypasses temporal smoothing and directly chooses the pod with the best spatial score:
\begin{equation}
p^*
=
\arg\min_p \mathrm{Score}_{\mathrm{spatial}}(p).
\label{eq:spatial_fallback}
\end{equation}
Therefore, degraded requests still preserve spatial balancing based on known disk size and predicted load intensity; only the unreliable temporal phase signal is removed.

\section{Details of Regex-Based Noise Filtering}
\label{app:noise_filter}

TIDAL uses a lightweight regex-based pre-filter to intercept provisioning metadata that are unlikely to contain recoverable application semantics before invoking the online student model. The filter is intentionally conservative: its goal is not to infer semantics, but only to reject \emph{obviously} low-information strings cheaply.

\noindent\textbf{Targeted low-information patterns.}
In production traces, the filter mainly targets three broad categories of metadata that rarely carry useful workload semantics:
\begin{itemize}[leftmargin=*,itemsep=1pt]
    \item \textbf{Identifier-like noise.} These are long mixed alphanumeric tokens, often separated by ``-'', ``\_'', or spaces, that resemble deployment-generated hashes, UUID fragments, or opaque internal identifiers.
    \item \textbf{Random-looking alphabetic fragments.} These include strings with highly unnatural orthographic structure, such as long consonant-heavy substrings or implausible letter combinations, which are unlikely to be human-assigned service names.
    \item \textbf{Default or degenerate names.} These include generic system-assigned labels (e.g., unnamed projects or default disk names) and strings that become nearly empty after removing obvious boilerplate tokens and separators.
\end{itemize}

\noindent\textbf{Conservative false-positive control.}
The filter is designed to reject only clear noise. To avoid suppressing meaningful metadata, it explicitly preserves common infrastructure and middleware terms such as database, cache, messaging, orchestration, and monitoring keywords. This allows names like \texttt{mysql-prod-01} or \texttt{kafka-cache-node} to proceed to semantic inference even when they contain digits or abbreviated tokens. More generally, if a string is not confidently recognized as low-information, it is passed to the student model rather than filtered out.

\noindent\textbf{Role in the pipeline.}
Requests intercepted by this filter do not enter semantic inference. Instead, they are routed directly to the confidence-based degradation path described in §\ref{sec:robustness}, where TIDAL falls back to intensity-aware spatial balancing. The filter therefore serves two purposes simultaneously: it reduces unnecessary model invocations and prevents obviously non-semantic identifiers from perturbing phase-aware placement.


\end{document}